\input harvmac.tex



\def\unlockat{\catcode`\@=11}
\def\lockat{\catcode`\@=12}

\unlockat

\def\newsec#1{\global\advance\secno by1\message{(\the\secno. #1)}
\global\subsecno=0\global\subsubsecno=0\eqnres@t\noindent
{\bf\the\secno. #1}
\writetoca{{\secsym} {#1}}\par\nobreak\medskip\nobreak}
\global\newcount\subsecno \global\subsecno=0
\def\subsec#1{\global\advance\subsecno
by1\message{(\secsym\the\subsecno. #1)}
\ifnum\lastpenalty>9000\else\bigbreak\fi\global\subsubsecno=0
\noindent{\it\secsym\the\subsecno. #1}
\writetoca{\string\quad {\secsym\the\subsecno.} {#1}}
\par\nobreak\medskip\nobreak}
\global\newcount\subsubsecno \global\subsubsecno=0
\def\subsubsec#1{\global\advance\subsubsecno by1
\message{(\secsym\the\subsecno.\the\subsubsecno. #1)}
\ifnum\lastpenalty>9000\else\bigbreak\fi
\noindent\quad{\secsym\the\subsecno.\the\subsubsecno.}{#1}
\writetoca{\string\qquad{\secsym\the\subsecno.\the\subsubsecno.}{#1}}
\par\nobreak\medskip\nobreak}

\def\subsubseclab#1{\DefWarn#1\xdef
#1{\noexpand\hyperref{}{subsubsection}%
{\secsym\the\subsecno.\the\subsubsecno}%
{\secsym\the\subsecno.\the\subsubsecno}}%
\writedef{#1\leftbracket#1}\wrlabeL{#1=#1}}
\lockat

\def\IL{\relax{\rm I\kern-.18em L}}
\def\IH{\relax{\rm I\kern-.18em H}}
\def\IR{\relax{\rm I\kern-.18em R}}
\def\IC{\relax\hbox{$\inbar\kern-.3em{\rm C}$}}
\def\IZ{\relax\ifmmode\mathchoice
{\hbox{\cmss Z\kern-.4em Z}}{\hbox{\cmss Z\kern-.4em Z}}
{\lower.9pt\hbox{\cmsss Z\kern-.4em Z}}
{\lower1.2pt\hbox{\cmsss Z\kern-.4em Z}}\else{\cmss Z\kern-.4em
Z}\fi}
\def\CM {{\cal M}}
\def\CN {{\cal N}}
\def\CR {{\cal R}}

\def\CO {{\cal O}}

\def\CH {{\cal H}}

\def\CM {{\cal M}}
\def\CN {{\cal N}}

\def\CO {{\cal O}}

\font\manual=manfnt \def\dbend{\lower3.5pt\hbox{\manual\char127}}

\def\IZ{\relax\ifmmode\mathchoice
{\hbox{\cmss Z\kern-.4em Z}}{\hbox{\cmss Z\kern-.4em Z}}
{\lower.9pt\hbox{\cmsss Z\kern-.4em Z}}
{\lower1.2pt\hbox{\cmsss Z\kern-.4em Z}}\else{\cmss Z\kern-.4em
Z}\fi}
\def\half {{1\over 2}}

\def\CM {{\cal M}}
\def\CN {{\cal N}}

\def\CO {{\cal O}}


\def\IZ{\relax\ifmmode\mathchoice
{\hbox{\cmss Z\kern-.4em Z}}{\hbox{\cmss Z\kern-.4em Z}}
{\lower.9pt\hbox{\cmsss Z\kern-.4em Z}}
{\lower1.2pt\hbox{\cmsss Z\kern-.4em Z}}\else{\cmss Z\kern-.4em
Z}\fi}
\def\IB{\relax{\rm I\kern-.18em B}}
\def\IC{{\relax\hbox{$\inbar\kern-.3em{\rm C}$}}}
\def\ID{\relax{\rm I\kern-.18em D}}
\def\IE{\relax{\rm I\kern-.18em E}}
\def\IF{\relax{\rm I\kern-.18em F}}
\def\IG{\relax\hbox{$\inbar\kern-.3em{\rm G}$}}
\def\IGa{\relax\hbox{${\rm I}\kern-.18em\Gamma$}}
\def\IH{\relax{\rm I\kern-.18em H}}
\def\II{\relax{\rm I\kern-.18em I}}
\def\IK{\relax{\rm I\kern-.18em K}}
\def\IP{\relax{\rm I\kern-.18em P}}

\def\inbar{\,\vrule height1.5ex width.4pt depth0pt}

\def\mod{{\rm mod}}

\font\cmss=cmss10 \font\cmsss=cmss10 at 7pt
\def\IR{\relax{\rm I\kern-.18em R}}


\def\boxit#1{\vbox{\hrule\hbox{\vrule\kern8pt
\vbox{\hbox{\kern8pt}\hbox{\vbox{#1}}\hbox{\kern8pt}}
\kern8pt\vrule}\hrule}}
\def\mathboxit#1{\vbox{\hrule\hbox{\vrule\kern8pt\vbox{\kern8pt
\hbox{$\displaystyle #1$}\kern8pt}\kern8pt\vrule}\hrule}}


\def\inbar{\,\vrule height1.5ex width.4pt depth0pt}

\font\cmss=cmss10 \font\cmsss=cmss10 at 7pt
\def\IR{\relax{\rm I\kern-.18em R}}


\def\a1{{\cal A}^{1,1}}


%
\lref\blzh{A. Belavin, V. Zakharov, ``Yang-Mills

Equations as inverse scattering problem''

Phys. Lett. B73, (1978) 53}
\lref\afs{Alekseev, Faddeev, Shatashvili,  }
\lref\mickelsson{ J. Mickelsson, ``Kac-Moody groups, topology of

the Dirac  determinant bundle and fermionization,''
Commun. Math. Phys. {\bf 110}(1987)173.}
\lref\bost{L. Alvarez-Gaume, J.B. Bost , G. Moore, P. Nelson, C.
Vafa,
``Bosonization on higher genus Riemann surfaces,''
Commun.Math.Phys.112:503,1987}
\lref\agmv{L. Alvarez-Gaum\'e,
C. Gomez, G. Moore,
and C. Vafa, ``Strings in the Operator Formalism,''
Nucl. Phys. {\bf 303}(1988)455}
\lref\atiyah{M. Atiyah, ``Green's Functions for
Self-Dual Four-Manifolds,'' Adv. Math. Suppl.
{\bf 7A} (1981)129}

\lref\AHS{M.~ Atiyah, N.~ Hitchin and I.~ Singer, ``Self-Duality in
Four-Dimensional
Riemannian Geometry", Proc. Royal Soc. (London) {\bf A362} (1978)
425-461.}
\lref\fmlies{M. F. Atiyah and I. M. Singer,
``The index of elliptic operators IV,'' Ann. Math. {\bf 93}(1968)119}
\lref\bagger{E. Witten and J. Bagger, Phys. Lett.
{\bf 115B}(1982)202}
\lref\banks{T. Banks, ``Vertex Operators in 2D Dimensions,''
hep-th/9503145   }
\lref\berk{N. Berkovits, ``Super-Poincare Invariant Superstring Field
Theory''
hep-th/9503099 }
\lref\biquard{O. Biquard, ``Sur les fibr\'es paraboliques
sur une surface complexe,'' to appear in J. Lond. Math.
Soc.}
\lref\bjsv{hep-th/9501096,
Topological Reduction of 4D SYM to 2D $\sigma$--Models,
 M. Bershadsky, A. Johansen, V. Sadov and C. Vafa }
\lref\BlThlgt{M.~ Blau and G.~ Thompson, ``Lectures on 2d Gauge
Theories: Topological Aspects and Path
Integral Techniques", Presented at the
Summer School in Hogh Energy Physics and
Cosmology, Trieste, Italy, 14 Jun - 30 Jul
1993, hep-th/9310144.}
\lref\bpz{A.A. Belavin, A.M. Polyakov, A.B. Zamolodchikov,
``Infinite conformal symmetry in two-dimensional quantum
field theory,'' Nucl.Phys.B241:333,1984}
\lref\braam{P.J. Braam, A. Maciocia, and A. Todorov,
``Instanton moduli as a novel map from tori to
K3-surfaces,'' Inven. Math. {\bf 108} (1992) 419}
\lref\cllnhrvy{Callan and Harvey, Nucl
Phys. {\bf B250}(1985)427}
\lref\CMR{ For a review, see
S. Cordes, G. Moore, and S. Ramgoolam,
`` Lectures on 2D Yang Mills theory, Equivariant
Cohomology, and Topological String Theory,''
Lectures presented at the 1994 Les Houches Summer School
 ``Fluctuating Geometries in Statistical Mechanics and Field
Theory.''
and at the Trieste 1994 Spring school on superstrings.
hep-th/9411210, or see http://xxx.lanl.gov/lh94}
\lref\devchand{Ch. Devchand and V. Ogievetsky,
``Four dimensional integrable theories,'' hep-th/9410147}
\lref\devchandi{
Ch. Devchand and A.N. Leznov,
``B \"acklund transformation for supersymmetric self-dual theories
for
semisimple
gauge groups and a hierarchy of $A_1$ solutions,'' hep-th/9301098,
Commun. Math. Phys. {\bf 160} (1994) 551}

\lref\elitzur{S. Elitzur, G. Moore,
A. Schwimmer, and N. Seiberg,
``Remarks on the Canonical Quantization of the Chern-Simons-
Witten Theory,'' Nucl. Phys. {\bf B326}(1989)108;
G. Moore and N. Seiberg,
``Lectures on Rational Conformal Field Theory,''
, in {\it Strings'89},Proceedings
of the Trieste Spring School on Superstrings,
3-14 April 1989, M. Green, et. al. Eds. World
Scientific, 1990}
\lref\etingof{P.I. Etingof and I.B. Frenkel,
``Central Extensions of Current Groups in
Two Dimensions,'' Commun. Math.
Phys. {\bf 165}(1994) 429}

\lref\evans{M. Evans, F. G\"ursey, V. Ogievetsky,
``From 2D conformal to 4D self-dual theories:
Quaternionic analyticity,''
Phys. Rev. {\bf D47}(1993)3496}
\lref\fs{L. Faddeev and S. Shatashvili, Theor. Math. Fiz., 60 (1984)
206, L. Faddeev, Phys. Lett. B145 (1984) 81, J. Mickelsson, CMP, 97
(1985) 361,
A. Reiman, M. Semenov-Tian-Shansky, L. Faddeev, Funct. Anal. and
Appl.,vol.18,No.4(1984)319}
\lref\fz{I. Frenkel, I. Singer, unpublished.}

\lref\fk{I. Frenkel and B. Khesin, ``Four dimensional
realization of two dimensional current groups,'' Yale
preprint, July 1995.}
\lref\galperin{A. Galperin, E. Ivanov, V. Ogievetsky,
E. Sokatchev, Ann. Phys. {\bf 185}(1988) 1}
\lref\gwdzki{K. Gawedzki, ``Topological Actions in Two-Dimensional
Quantum Field Theories,'' in {\it Nonperturbative
Quantum Field Theory}, G. 't Hooft, A. Jaffe, et. al. , eds. ,
Plenum 1988}
\lref\gmps{A. Gerasimov, A. Morozov, M. Olshanetskii,
 A. Marshakov, S. Shatashvili ,``
Wess-Zumino-Witten model as a theory of
free fields,'' Int. J. Mod. Phys. A5 (1990) 2495-2589
 }
\lref\gerasimov{A. Gerasimov, ``Localization in
GWZW and Verlinde formula,'' hepth/9305090}
\lref\ginzburg{V. Ginzburg, M. Kapranov, and E. Vasserot,
``Langlands Dualtiy for Surfaces,'' IAS preprint}
\lref\giveon{hep-th/9502057,
 S-Duality in N=4 Yang-Mills Theories with General Gauge Groups,
 Luciano Girardello, Amit Giveon, Massimo Porrati, and Alberto
Zaffaroni
}

\lref\gottsh{L. Gottsche, Math. Ann. 286 (1990)193}
\lref\gothuy{L. G\"ottsche and D. Huybrechts,
``Hodge numbers of moduli spaces of stable
bundles on $K3$ surfaces,'' alg-geom/9408001}
\lref\GrHa{P.~ Griffiths and J.~ Harris, {\it Principles of
Algebraic
geometry},
p. 445, J.Wiley and Sons, 1978. }

\lref\hitchin{N. Hitchin, ``Polygons and gravitons,''
Math. Proc. Camb. Phil. Soc, (1979){\bf 85} 465}

\lref\hklr{Hitchin, Karlhede, Lindstrom, and Rocek,
``Hyperkahler metrics and supersymmetry,''
Commun. Math. Phys. {\bf 108}(1987)535}
\lref\hirz{F. Hirzebruch and T. Hofer, Math. Ann. 286 (1990)255}
\lref\hms{hep-th/9501022,
 Reducing $S$- duality to $T$- duality, J. A. Harvey, G. Moore and A.
Strominger}
\lref\johansen{A. Johansen, ``Infinite Conformal
Algebras in Supersymmetric Theories on
Four Manifolds,'' hep-th/9407109}
\lref\kronheimer{P. Kronheimer, ``The construction of ALE spaces as
hyper-kahler quotients,'' J. Diff. Geom. {\bf 28}1989)665}
\lref\kricm{P. Kronheimer, ``Embedded surfaces in
4-manifolds,'' Proc. Int. Cong. of
Math. (Kyoto 1990) ed. I. Satake, Tokyo, 1991}

\lref\KN{Kronheimer and Nakajima,  ``Yang-Mills instantons
on ALE gravitational instantons,''  Math. Ann.
{\bf 288}(1990)263}
\lref\krmw{P. Kronheimer and T. Mrowka,
``Gauge theories for embedded surfaces I,''
Topology {\bf 32} (1993) 773,
``Gauge theories for embedded surfaces II,''
preprint.}

\lref\lmns{A. Losev, G. Moore, N. Nekrasov, and
S. Shatashvili,  ``Four-Dimensional Avatars of
Two-Dimensional RCFT,''  hep-th/9509151.}

\lref\maciocia{A. Maciocia, ``Metrics on the moduli
spaces of instantons over Euclidean 4-Space,''
Commun. Math. Phys. {\bf 135}(1991) , 467}
\lref\mick{I. Mickellson, CMP, 97 (1985) 361.}

\lref\milnor{J. Milnor, ``A unique decomposition
theorem for 3-manifolds,'' Amer. Jour. Math, (1961) 1}
\lref\taming{G. Moore and N. Seiberg,
``Taming the conformal zoo,'' Phys. Lett.
{\bf 220 B} (1989) 422}
\lref\nair{V.P.Nair, ``K\"ahler-Chern-Simons Theory'', hep-th/9110042
}
\lref\ns{V.P. Nair and Jeremy Schiff,
``Kahler Chern Simons theory and symmetries of
antiselfdual equations'' Nucl.Phys.B371:329-352,1992;
``A Kahler Chern-Simons theory and quantization of the
moduli of antiselfdual instantons,''
Phys.Lett.B246:423-429,1990,
``Topological gauge theory and twistors,''
Phys.Lett.B233:343,1989}
\lref\nakajima{H. Nakajima, ``Homology of moduli
spaces of instantons on ALE Spaces. I'' J. Diff. Geom.
{\bf 40}(1990) 105; ``Instantons on ALE spaces,
quiver varieties, and Kac-Moody algebras,'' preprint,
``Gauge theory on resolutions of simple singularities
and affine Lie algebras,'' preprint.}
\lref\nakheis{H.Nakajima, ``Heisenberg algebra and Hilbert schemes of
points on
projective surfaces ,'' alg-geom/9507012}
\lref\ogvf{H. Ooguri and C. Vafa, ``Self-Duality
and $N=2$ String Magic,'' Mod.Phys.Lett. {\bf A5} (1990) 1389-1398;
``Geometry
of$N=2$ Strings,'' Nucl.Phys. {\bf B361}  (1991) 469-518.}
\lref\park{J.-S. Park, ``Holomorphic Yang-Mills theory on compact
Kahler
manifolds,'' hep-th/9305095; Nucl. Phys. {\bf B423} (1994) 559;
J.-S.~ Park, ``$N=2$ Topological Yang-Mills Theory on Compact
K\"ahler
Surfaces", Commun. Math, Phys. {\bf 163} (1994) 113;
J.-S.~ Park, ``$N=2$ Topological Yang-Mills Theories and Donaldson
Polynomials", hep-th/9404009}
\lref\parki{S. Hyun and J.-S. Park,
``Holomorphic Yang-Mills Theory and Variation
of the Donaldson Invariants,'' hep-th/9503036}
\lref\pohl{Pohlmeyer, Commun.
Math. Phys. {\bf 72}(1980)37}
\lref\pwf{A.M. Polyakov and P.B. Wiegmann,
Phys. Lett. {\bf B131}(1983)121}
\lref\prseg{A.~Pressley and G.~Segal, "Loop Groups", Oxford Clarendon
Press, 1986}
\lref\rade{J. Rade, ``Singular Yang-Mills fields. Local
theory I. '' J. reine ang. Math. , {\bf 452}(1994)111; {\it ibid}
{\bf 456}(1994)197; ``Singular Yang-Mills
fields-global theory,'' Intl. J. of Math. {\bf 5}(1994)491.}
\lref\segal{G. Segal, The definition of CFT}
\lref\seiberg{hep-th/9407087,
Monopole Condensation, And Confinement In $N=2$ Supersymmetric
Yang-Mills
Theory, N. Seiberg and E. Witten;
hep-th/9408013,  Nathan Seiberg;
hep-th/9408099,
Monopoles, Duality and Chiral Symmetry Breaking in
N=2 Supersymmetric QCD, N. Seiberg and E. Witten;
hep-th/9408155,
Phases of N=1 supersymmetric gauge theories in four dimensions, K.
Intriligator
and N. Seiberg; hep-ph/9410203,
Proposal for a Simple Model of Dynamical SUSY Breaking, by K.
Intriligator, N.
Seiberg, and S. H. Shenker;
hep-th/9411149,
 Electric-Magnetic Duality in Supersymmetric Non-Abelian Gauge
Theories,
 N. Seiberg; hep-th/9503179 Duality, Monopoles, Dyons, Confinement
and Oblique
Confinement in Supersymmetric $SO(N_c)$ Gauge Theories,
K. Intriligator and N. Seiberg}
\lref\sen{A. Sen,
hep-th/9402032, Dyon-Monopole bound states, selfdual harmonic
forms on the multimonopole moduli space and $SL(2,Z)$
invariance,'' }
\lref\shatashi{S. Shatashvili,
Theor. and Math. Physics, 71, 1987, p. 366}
\lref\thooft{G. 't Hooft , ``A property of electric and
magnetic flux in nonabelian gauge theories,''
Nucl.Phys.B153:141,1979}
\lref\vafa{C. Vafa, ``Conformal theories and punctured
surfaces,'' Phys.Lett.199B:195,1987 }
\lref\VaWi{C.~ Vafa and E.~ Witten, ``A Strong Coupling Test of
$S$-Duality",
hep-th/9408074.}
\lref\vrlsq{E. Verlinde and H. Verlinde,
``Conformal Field Theory and Geometric Quantization,''
in {\it Strings'89},Proceedings
of the Trieste Spring School on Superstrings,
3-14 April 1989, M. Green, et. al. Eds. World
Scientific, 1990}
\lref\mwxllvrld{E. Verlinde, ``Global Aspects of
Electric-Magnetic Duality,'' hep-th/9506011}
\lref\wrdhd{R. Ward, Nucl. Phys. {\bf B236}(1984)381}
\lref\ward{Ward and Wells, {\it Twistor Geometry and
Field Theory}, CUP }
\lref\wittenwzw{E. Witten, ``Nonabelian bosonization in
two dimensions,'' Commun. Math. Phys. {\bf 92} (1984)455 }
\lref\grssmm{E. Witten, ``Quantum field theory,
grassmannians and algebraic curves,'' Commun.Math.Phys.113:529,1988}
\lref\wittjones{E. Witten, ``Quantum field theory and the Jones
polynomial,'' Commun.  Math. Phys.}
\lref\wittentft{E.~ Witten, ``Topological Quantum Field Theory",
Commun. Math. Phys. {\bf 117} (1988) 353.}
\lref\Witdgt{ E.~ Witten, ``On Quantum gauge theories in two
dimensions,''
Commun. Math. Phys. {\bf  141}  (1991) 153.}
\lref\Witfeb{E.~ Witten, ``Supersymmetric Yang-Mills Theory On A
Four-Manifold,'' J. Math. Phys. {\bf 35} (1994) 5101.}
\lref\Witr{E.~ Witten, ``Introduction to Cohomological Field
Theories",
Lectures at Workshop on Topological Methods in Physics, Trieste,
Italy,
Jun 11-25, 1990, Int. J. Mod. Phys. {\bf A6} (1991) 2775.}
\lref\wittabl{E. Witten,  ``On S-Duality in Abelian Gauge Theory,''
hep-th/9505186}
\lref\nov{ S. Novikov,"The Hamiltonian
formalism and many-valued analogue of  Morse theory",
Russian Math.Surveys 37:5(1982),1-56;
E. Witten, ``Global Aspects of Current Algebra,''
Nucl.Phys.B223:422,1983}
\lref\faddeevlmp{ L. D. Faddeev, ``Some Comments on Many Dimensional
Solitons'',
Lett. Math. Phys., 1 (1976) 289-293.}

\lref\cvi{S. Cecotti and C. Vafa, ``On classification
of N=2 supersymmetric theories,''
hep-th/????}

\lref\hmi{J. Harvey and G. Moore, ``On the algebras of
BPS states,''}

\lref\intfen{K. Intriligator and P. Fendley, ``Scattering
and thermodynamics in integrable $N=2$ theories,''
hep-th/9202011}

\lref\katzvafa{S. Katz and C. Vafa, ``Geometric
engineering of $N=1$ quantum field theories,''
hep-th/9611090.}

\lref\zas{E. Zaslow, ``Solitons and helices: the search
for a math-physics bridge,'' CMP {\bf 175}(1996)337}


\lref\chs{C. Callan, J. Harvey and A. Strominger, ``Worldbrane 
actions for string solitons,'' Nucl. Phys.
{\bf B367} (1991) 60; ``Supersymmetric string solitons,'' 
hep-th/9112030}

\lref\christ{N.H. Christ, E.J. Weinberg, and
N.K. Stanton, ``General self-dual Yang-Mills
solutions,''  Phys. Rev. {\bf D18}(1978)2013}
\lref\corrigan{E. Corrigan and P. Goddard,
``Construction of instantons and monopole
solutions and reciprocity,'' Ann. Phys. {\bf 154} (1984)253}

\lref\diac{ E. Diaconescu, }

\lref\dgmr{M. Douglas and G. Moore,
``D-branes, Quivers and  ALE Instantons,''
hep-th/9603167}

\lref\dnld{S. Donaldson, ``Anti self-dual Yang-Mills
connections over complex  algebraic surfaces and stable
vector bundles,'' Proc. Lond. Math. Soc,
{\bf 50} (1985)1}

\lref\DoKro{S.K.~ Donaldson and P.B.~ Kronheimer,
{\it The Geometry of Four-Manifolds},
Clarendon Press, Oxford, 1990.}
\lref\donii{
S. Donaldson, Duke Math. J. , {\bf 54} (1987) 231. }

\lref\garmi{H. Garland and M.K. Murray,
``Kac-Moody Monopoles and Periodic Instantons,''
CMP 120 335 (1988)}
\lref\garmii{H. Garland and M.K. Murray,
``Why instantons are monopoles,''
CMP 121 85 (1988)}

\lref\gpy{Gross, Pisarski, Yaffe, Rev. Mod. Phys. {\bf 53}1 1981}

\lref\hanwitt{Hanany and Witten}

\lref\hurti{J. Hurtubise and M.K. Murray,
``On the construction of monopoles for the
classical groups,'' CMP 122 35 }
\lref\hurtii{J. Hurtubise and M.K. Murray,
CMP 133 487}
\lref\hurtiii{
J. Hurtubise CMP 120 613}

\lref\kn{P.B. Kronheimer and H. Nakajima, ``Yang-Mills instantons
on ALE gravitational instantons,'' Math. Ann. {\bf 288} (1990) 263.}
\lref\witten{E. Witten, ``Small Instantons in String Theory,'' hep-th/9511030.}


\lref\bds{T. Banks, M.R. Douglas, and N. Seiberg, ``Probing
F-theory with branes,'' hep-th/9605199}
\lref\bfss{T. Banks, W. Fischler, S. Shenker, and L. Susskind,
``M theory as a matrix model: a conjecture,''
hep-th/9610043}
\lref\brs{M. Berkooz, M. Rozali, and N. Seiberg, ``Matrix 
description of M-theory on $T^4$ and $T^5$, hep-th/9704089}
\lref\cj{E. Cremmer and B. Julia, ``The $SO(8)$ supergravity''
Nuc. Phys. {\bf B159}(1979)141}
\lref\DVVsix{R. Dijkgraaf, E. Verlinde, and H. Verlinde, ``BPS spectrum 
of the five-brane and black hole entropy,'' hep-th/9603126;
``BPS quantization 
of the five-brane,'' hep-th/9604055; }
\lref\DVVsev{R. Dijkgraaf, E. Verlinde, and H. Verlinde, ``5D black holes and 
matrix strings,'' hep-th/9704018} 
\lref\dkps{M. Douglas, D. Kabat, P. Pouliot, and 
S. Shenker, ``D-branes and short distances in string 
theory,'' Nucl. Phys. {\bf B485}(1997)85;hep-th/9608024}
\lref\dps{M. Douglas, J. Polchinski, and A. Strominger, 
``Probing five-dimensional black holes with D-branes,'' 
hep-th/9703031}
\lref\egk{S. Elitzur, A. Giveon, D. Kutasov, and 
E. Rabinovici, ``Algebraic aspects of matrix theory on 
$T^d$, hep-th/9707217 }
\lref\hull{C. Hull and P. Townsend, ``Unity of superstring dualities,'' 
hep-th/9410167 }
\lref\vfmr{D. Morrison and C. Vafa, ``Compactifications of F-theory on 
Calabi-Yau threefolds -I,'' hep-th/9602114;``Compactifications of F-theory on 
Calabi-Yau threefolds -II,'' hep-th/9603161} 
\lref\nnfive{N. Nekrasov, ``Five-dimensional gauge theories and 
relativistic integrable systems,'' hep-th/9609219}
\lref\polchinski{S. Chaudhuri, C. Johnson, and J. Polchinski,
``Notes on D-branes,'' hep-th/9602052; J. Polchinski,
``TASI Lectures on D-branes,'' hep-th/9611050}
\lref\samson{S. Shatashvili, ``Open strings, closed strings, and 
matrix theory,'' to appear}
\lref\schwarz{J. Schwarz, ``Self-dual superstring in six dimensions,'' 
hep-th/9604171}
\lref\seibir{N. Seiberg, ``IR dynamics on branes and 
space-time geometry,'' hep-th/9606017}
\lref\sls{N. Seiberg, ``New theories in six dimensions and matrix description
of M-theory on $T^5$ and $T^5/Z_2$,'' hep-th/9705221}
\lref\seibhiggs{O. Aharony, M. Berkooz, S. Kachru, N. Seiberg, and 
E. Silverstein, ``Matrix description of interacting theories in six 
dimensions,'' hep-th/9707079}
\lref\seibcoul{D.-E. Diaconescu and N. Seiberg, ``The coulomb branch 
of (4,4) supersymmetric field theories in two dimensions,'' hep-th/9707158}
\lref\senpillow{A. Sen, ``F theory and orientifolds,'' hep-th/9605150 }
\lref\sentn{A. Sen, 9705212; ``Dynamics of multiplet Kaluza-Klein monopoles in 
M- and string theory,'' hep-th/9707042; ``A note on enhanced gauge symmetries in 
M- and string theory,'' hep-th/9707123}
\lref\stromopen{A. Strominger, ``Open p-branes,'' hep-th/9512059 }
\lref\stromvafa{A. Strominger and C. Vafa, ``Microscopic origin of the 
Beckenstein-Hawking entropy,'' hep-th/9601029} 
\lref\townsend{P. Townsend, ``The eleven dimensional supermembrane
revisited,'' hep-th/9501068}
\lref\vafa{C. Vafa, ``Evidence for F-theory,'' hep-th/9602022}
\lref\WittVar{E. Witten, ``String theory in various dimensions,'' 
hep-th/9503124}
\lref\WittCommt{E. Witten, ``Some comments on string dynamics,'' hep-th/9507121}
\lref\WittHiggs{E. Witten, ``On the conformal field theory of the 
Higgs Branch,'' hep-th/9707093}
\lref\WittQCD{E. Witten, ``Branes and the dynamics of QCD,'' hep-th/9706109}

\Title{ \vbox{\baselineskip12pt\hbox{hep-th/9707250}
 \hbox{ITEP-35/97}
\hbox{YCTP-P13-97}}}
{\vbox{
\centerline{ M \& m's }}}
\vskip 1cm
\medskip
\vskip 0.5cm
\centerline{{
Andrei Losev \footnote{*}{Institute of Theoretical and Experimental
Physics,
117259, Moscow, Russia.}, Gregory Moore,  and Samson L. Shatashvili 
\footnote{**}{On
leave of
absence from St. Petersburg Steklov Mathematical Institute, St.
Petersburg,
Russia.}}}
\vskip 0.5cm
\centerline{ Dept.\ of Physics, Yale University,}
\centerline{P.O. Box 208120, New Haven, CT  06520}
\vskip 0.5cm

\medskip
\noindent
\bigskip
\bigskip
\noindent
We consider $M$ theory 5-branes 
with compact transverse dimensions. 
In certain limits the theory on the 5-brane decouples 
and defines ``little string theories'' in $5+1$ dimensions.  
We show that the  familiar structure of 
$IIA/IIB,M,F-$ theory
in $10,11,12$ 
dimensions respectively has a perfect 
parallel in a theory of  
strings and membranes in $6,7,8$ dimensions. 
We call these theories $a/b,m,f$ theories.    
They have a coupling constant but no gravity. 
This construction clarifies some mysteries in 
$F$-theory and leads to several speculations 
about the phase structure of $M$ theory. 

\Date{July 30, 1997}

\newsec{Introduction}

In the past year six-dimensional string theories have 
come under intense scrutiny \DVVsix\schwarz\DVVsev
\brs\sls\seibhiggs\WittHiggs\seibcoul. 
The new six-dimensional string theories
are interesting in their own right and have many 
potential applications. They are both similar to 
and different from ten-dimensional strings. 
They are similar in that they describe Lorentz-invariant 
interacting strings with finite tension and exhibit duality 
symmetries. They are radically different because they 
are six-dimensional and 
contain no graviton. It is possible that they they contain, 
in some sense, many fewer degrees of freedom. 
For all these reasons 
we will refer to these theories as ``little string theories.''
\foot{The authors of \DVVsix\ have used the term 
``microstring theories.''}

In this note we would like to add two elements to the 
discussion on little string theories: 1.) The existence
of the six-dimensional strings suggests the existence 
of a seven-dimensional superPoincar\'e invariant theory 
whose low energy limit is 6+1 supersymmetric Yang-Mills
theory (SYM). We call this theory 
$m$-theory. 2.) The little string theories have 
(little) $d$-branes and a dimensionless parameter which 
we interpret as a coupling constant. 

The construction of the theories proceeds by taking limits
of the $11,10$ dimensional theories in the presence of 
$k$ 5-branes,  6-branes, or a multi-Taub-NUT geometry, 
depending on the duality frame. The limiting theory is 
the theory of strings ``captured'' by the 5-branes or
by the centers of the Taub-NUT space. The existence of 
these captured strings, while surprising, is demanded 
by $U$-duality. The precise dynamics of 
the capturing mechanism has been recently studied 
  in terms of the 
decoupling of Coulomb and Higgs branches of a $D1$- or 
$D2$-brane probe \seibhiggs\WittHiggs\seibcoul. 

The spectrum of excited states in $m,a,b$ theories 
parallels that in $11,10$ dimensions. The $m_k$ theory 
associated to $k$ parallel $D6$ branes
contains no strings but does contain $k^2$ interacting 
membranes with a $U(k)$ symmetry. The low energy limit 
is $U(k)$ $6+1$ SYM and does not contain the graviton 
multiplet. Similarly, the 
$a_k$ theory contains $k^2$ interacting strings, 
$d0$ and $d2$ branes, while the $b_k$ theory contains 
$k^2$ $SL(2,\IZ)$ multiplets of strings. The low 
energy limit of this theory is the mysterious nonabelian 
tensor multiplet theory with $U(k)$ gauge invariance whose
existence was predicted in \WittCommt\stromopen.

The paper is  organized as follows. 
In   the next section we outline the main
ideas leading to  $m$-theory. 
A detailed discussion of the limits which should 
define the little theories is given in section 3. 
The duality symmetries of the little strings 
are discussed in section 4. Section 5 contains 
some remarks on BPS states, and section 6 makes some 
preliminary remarks on the phenomenon of 
captured strings.  

During the course of this work several papers have 
appeared with substantial overlap with our work.
This includes \sls\seibhiggs\WittHiggs\seibcoul\sentn. 
We hope that the  two points mentioned above 
are novel (and correct!).

\newsec{From $M$-theory to $m$-theory }

\subsec{Cast of characters}

We would like to search for a lower-dimensional 
analogue of $M$-theory. Our desiderata include 
Lorentz invariance, existence of extended objects, 
duality symmetry, and geometrically defined interactions. 
Moreover, in searching for a minimal or simplest 
version of lower-dimensional $M$ theory 
we ask that the theory have no gravity, since a 
theory with gravity is likely to end up being a 
well-known phase of $M$-theory.

A natural candidate for such a lower dimensional theory 
without gravity is a theory on a brane. However, in general, 
the notion of ``the theory on the brane'' is not well-defined. 
The modes of excitation of the brane couple to the bulk modes. 
For example, consider scattering of particles within the brane. 
These will inevitably lead to gravitational radiation which 
will radiate off the brane into the bulk spacetime in which 
the brane is embedded. 
The amplitude of the radiation is suppressed by 
the Planck scale $E_G=M_{\rm Planck}$ of the bulk noncompactified theory. 
In low energy processes
($E<<E_G$) gravitational radiation can be neglected.
 Thus we search for interacting theories 
that have extended objects (p-branes) with
tensions $T_p$ much smaller than
the scale of emmission of gravity: 
\eqn\zero{ E_p=(T_p)^{1/(p+1)}<< E_G.}
The presence of such extended objects is expected to
smooth the ultraviolet divergences (as ordinary strings do)
at the scale  $E_p$,
and produce a finite theory (when we finally take the limit
$E_p/E_G \rightarrow 0$ \foot{This procedure is quite
analogous to the search for finite field theories. 
One  starts with a regularized theory (by adding 
massive particles or by embedding into a string theory) and then
(if there are no divergences) one takes a limit in  which
mass of the regulator goes to infinity. }.

In our search we should try to find the maximally
general theories in this class. Thus, we look for the 
maximal dimension brane. An important 
restriction on our search is that the brane produce a 
well-defined string
background.
Thus, the dimension of noncompact space
transverse to the brane has to be 3 or larger. 
The reason is that 
 the fields in the bulk must 
solve the Laplace equation and also must
decay to the flat background at infinity. If there are
two noncompact dimensions then there is an 
unacceptable logarithmic growth at infinity. 
Thus, we conclude that the maximal possible 
spatial dimension 
of the brane is 
$6$ in type II theories  and  $7$ in $M$-theory.
These considerations isolate the $6$-brane of type IIA 
theory as a distinguished theory. 
\foot{These and other problems that one 
encounters in higher dimensions, including 10, will 
be discussed in \samson.} We will denote 
a IIA string background
with a collection of $k$ parallel $D6$ branes as $(D6, g , T_{1A} )_k$,
 where $g$ is the string coupling constant
and $T_{1A}$ is the tension of the string. 

Using compactification and dualities one can relate this 
distinguished background to several other interesting backgrounds. 
In particular, if the $D6$ brane wraps a longitudinal circle  
then we can dualize to backgrounds with various 
$5$-branes inserted.  We will denote such compactifications as
$(NS5(A),R,g,T_{1A})_k$,  $(NS5(B),R,g,T_{1B})_k$  and  $(D5,R,g,T_{1B})_k$. 
$NS5$ denotes the solitonic 5-brane of \chs\ while $D5$ 
denotes the Dirichlet 
5-brane.
\foot{The solitonic 5-brane is often called the NS 5-brane because it is 
charged under the NS $B$-field. Surely there is a more 
rational terminology!} $R$ denotes the radius of a circle
transverse to the 5-brane. The corresponding $M$-theory 
background is denoted as $(M5, R_1,R_2, T_{M2})_k$ where $R_1,R_2$ are 
the radii of the transverse torus and the $T_{M2}$ is the 
tension of the $M$-theory 2-brane $T_{M2} = 1/\ell^3$, where
$\ell$ is the 
eleven-dimensional Planck length. 
   
\subsec{The Taub-NUT picture}

The $T$-dual along a circle transverse to a solitonic 5brane is a 
Taub-NUT space. Therefore, all 
of the above backgrounds can be related to backgrounds involving 
(multi)Taub-NUT gravitational 
instantons. 

A  multi-TN geometry  is defined by the 4-dimensional metric: 
\eqn\tngeom{
\eqalign{
ds^2 & = U^{-1} (dx^4 + \vec \omega\cdot \vec dx)^2 + 
U (\vec dx)^2  \cr
dU & = *_3 \vec \omega\cdot \vec dx\cr
U& = V^{-2}+ \half \sum_{i=1}^k {\ell \over \vert \vec x - \vec x_i\vert} \cr
0 & \leq x^4 \leq 2 \pi \ell \cr}
}
In $M$ theory such a geometry defines a $p=6$ brane while in 
type II theory such a geometry defines a $p=5$ brane.
(From the $M$-theory perspective the TN geometry is 
transverse to the circle defining the string limit.) 

The asymptotic Taub-NUT geometry  for $M$
theory is characterized by  one dimensionless parameter, 
the radius $V$ of the Hopf fiber in the squashed 3-sphere 
as measured in 11-dimensional Planck units.
The type II strings in a Taub-NUT geometry have
two dimensionless parameters: one is the string coupling constant $g$,
the other is the  radius of the Taub-NUT measured in
string   units. 
We  denote the (multi)Taub-NUT brane with the radius $R$
 in IIA, IIB and 11 dimensional phases as
$(TN5[A,R], g,T_{1A} )_k $, $(TN5[B,R],g,T_{1B})_k$ and $(TN6[M,R],T_{M2})_k$
respectively. The explicit dual relations between the
backgrounds are: 

\eqn\onea{ (D6,g,t)_k = \biggl(TN6\bigl[M, {g \over \sqrt{t}} \bigr],t^{3/2}/g\biggr)_k }

\eqn\oneb{\eqalign{(D5,R,g,t)_k & = 
\bigl( NS5(B),R,1/g,t/g\bigr)_k \cr 
& = \biggl(TN5[A,{g\over Rt} ],{1 \over R \sqrt{tg}} ,t/g\biggr)_k \cr }}

\eqn\onem{\eqalign{ (M5,R_1,R_2 ,T_{M2}=\ell^{-3} )_k & =(NS(A),R_2, 
(R_1/\ell)^{3/2} , R_1\ell^{-3} )_k  \cr
=&\biggl(TN5\bigl[B,{\ell^3 \over (R_1R_2)}\bigr], R_1/R_2,R_1 \ell^{-3}
\biggr)_k.\cr }}

The equality \onea\ follows from the promotion of 
$IIA$-theory to $M$-theory \townsend\WittVar. 
We have used $S$-duality in \oneb.

\subsec{Definition of the little theories}

It is quite instructive  to write the above 
theories in 
``Taub-NUT variables.''   In TN variables the above three 
backgrounds are expressed as: 

\eqn\m{ \biggl(TN5\bigl[M,V\ell \bigr] 
,T_{M2}=\ell^{-3} \biggr)_k = (D6, V^{3/2} , V\ell^{-2} )_k ,}

\eqn\twoa{\eqalign{\biggl(TN5\bigl[A,{V\over \sqrt{t}} \bigr],g ,t\biggr)_k & = 
\bigl( NS5(B), {1 \over V\sqrt{t} }, {g\over V} , t\bigr)_k  \cr
& =\bigl(D5, {1\over V\sqrt{t}}  , {V\over g}  , tV/g\bigr)_k \cr }}

\eqn\twob{\eqalign{\biggl(TN5\bigl[B,{V\over \sqrt{t}} \bigr],g ,t\biggr)_k &=
(NS5(A), {1 \over V\sqrt{t} } , {g\over V} ,t)_k \cr
& = \biggl(M5,{g \over V\sqrt{t} } , {1 \over V\sqrt{t} } , {t^{3/2}V\over g}\biggr)_k \cr}}

In the next section we 
compare the bulk Planck mass with the tensions
of extended objects on the branes in various duality frames,
using the relation
$ g^2 (M_{\rm Planck}^{(10)})^{8}
= (T_{1A,B})^4$ between the 10-dimensional Plank mass $M_{\rm Planck}^{(10)}$,
string coupling constant $g$, and string tension $T_{1A,B}$. 
The ratio of scales given by these tensions and
the Planck mass goes to zero if and only if
$V \rightarrow \infty$ 
(while keeping other parameters in the Taub-NUT picture fixed). 

{\it Therefore, we define the $m,a,b$ theories as
limits of theories \m, \twoa\ and \twob\ when $V \rightarrow \infty $, respectively. }

In the next several sections we will show that: 
 
\item{1.} 
The theories whose worldvolume is at the center of the Taub-NUT contains
all $p$-branes(for $p<3$) of the bulk theory,
i.e.  $m$-theory contains 2-branes (captured from 
the 2-branes  of the M-theory), 
$IIa$-theory contains membranes and strings, while 
$IIb$-theory contains an $SL(2,\IZ)$ multiplet of strings.
Moreover, $a$ and $b$ theories have $d$-branes and hence
are interacting string theories.

\item{2.}  Compactifying   $m$-theory on a circle gives  a-theory
and  the group of $u$-dualities of $m$- theory 
follows from the $T$ and $S$ dualities of   $M$-theory compactified
on the Taub-NUT.

\item{3.} The little theories have a rich BPS spectrum, but do not 
contain gravity.

>From 1,2,3 we see that the 
the brane defined by the center of a Taub-NUT geometry 
can be regarded as a ``universal
attractor'' of the branes  of the bulk theory.
Moreover, it attracts only those branes that have
codimension bigger than two within the brane.

\newsec{Snapshots of the $a/b,m,f$ theories }

In this section we give a more detailed 
description of the scales defined by the captured 
branes in various pictures of the 5brane.  
A key point in what follows is that captured strings 
are interacting. 
Captured fundamental
strings can end on captured $D$-branes. 
Thus the little string has $d$-branes. We 
postulate that the $d$ $p$-brane has tension 
$\sim {1 \over g} t^{(p+1)/2} $ where $t$ is 
the tension of a little string. This defines a 
dimensionless coupling $g$ for the little string
theory. 

Let us introduce some notation. 
We will consider compactification on a product of 
circles with a diagonal metric and with background 
antisymetric tensor fields put to zero. 
We denote directions by their
radii. Radii which are not explicitly written are assumed
to be infinite. If a background has a brane inserted
we put a bar over directions which are wrapped
by a brane. A subscript indicates multiplicity
or brane type. 
 No bar indicates that the direction is
not wrapped. Couplings and/or tensions needed to specify 
the theory are listed after the semicolon.  
We now discuss the same limit in several different duality 
frames, or pictures. 

\subsec{The tensormultiplet picture: Definition 1 of 
$IIb$-theory}

The easiest ``little string theory'' to define is IIb. 
We start with the system: 
\eqn\smsxi{
M:  (\overline{R_1,R_2,R_3,R_4,R_5})_{k,M5}, R_6,R_7 ; T_{M2} = \ell^{-3}
}
The 5-brane theory has two basic strings from the wrapped 
M2-branes of tensions
\eqn\twostrngs{
t_{b10} = R_6/\ell^3 \qquad\qquad t_{b01} = R_7/\ell^3
}
We will find that the limit in which 
gravity decouples involves shrinking the radii 
$R_6,R_7$ and hence we should compare energies 
to the 9-dimensional Planck scale. 
Thus, gravity decouples when we
perform experiments on the 5brane at energies 
\foot{Experiments performed at energies 
$E_{\rm expt}\leq \sqrt{t_{b10}}, \sqrt{t_{b01}}$ 
are described by the tensormultiplet field theory.}
\eqn\energies{
\sqrt{t_{b10}}, \sqrt{t_{b01}} \leq E_{\rm expt} \ll M_{\rm Planck}^{(9)}=
\bigl( {R_6 R_7 \over \ell^9}\bigr)^{1/7} }
Equivalently, we take $\ell \rightarrow 0$ holding 
the string tensions fixed: 
\eqn\EMM{
\eqalign{
R_6/\ell & = t_{b10} \ell^2 \rightarrow 0 \cr
R_7/\ell & = t_{b01} \ell^2 \rightarrow 0 \cr}
}
The definition of 
IIb theory depends on introducing a scale and 
a dimensionless coupling constant $g_b$, or 
equivalently, two string tensions $t_{b10}$ and 
$t_{b01}$. From its origin in the bulk
the little  $(01)$-string is a Dirichlet string for 
the little 
$(10)$-string so the coupling is $g_{b}= R_6/R_7$,
just as for the bulk $IIB$ string. 
 In summary: 
\eqn\deftwob{
\eqalign{
IIb: \quad R_1,R_2,R_3,R_4,R_5; 
& g_b=t_{b10}/t_{b01}, t_{1b} = t_{b10}
 \cr
\equiv \lim_{\ell \rightarrow 0} 
M:  (\overline{R_1,R_2,R_3,R_4,R_5})_{M5}, 
&
R_6=t_{b10} \ell^3 ,R_7= t_{b01} \ell^3 ; T_{M2} = \ell^{-3} \cr}
}

After transformation to the  picture of IIB compactified
on Taub-NUT we see that strings on the brane are transformed into
the $(1,0)$ and $(0,1)$ strings
of type IIB theory, captured by center of
Taub-NUT. The dimensionless parameters
$R_6/\ell,R_7/\ell $ expressed in
Taub-NUT units are equal to $1/(V^2 g_B)^{1/3}$ and
$(g_B/V)^{2/3}$ respectively.
Thus, we have shown that the limit of decoupling of gravity
decribed above really corresponds to large $V$ limit.

\subsec{The tensormultiplet picture: Definition 2 of 
$IIb$-theory}

Closely related to the above defintion of $IIb$ theory 
is the definition following from the NS5-brane 
transverse to a compact direction. 
Thus we consider the theory: 
\eqn\lttlebp{
IIA:  (\overline{R_1,R_2,R_3,R_4,R_5})_{k,NS},R_6 ; g_A , T_{A1}  
}

The definition of the little string in this 
duality frame is motivated by considering 
certain excitations on the 5brane. In particular 
there is a fundamental string of tension 
$t_{1b}=T_{1A}$
embedded in the NS5 brane.  
 In addition, the wrapped $D2$ brane can end on
the NS5 brane giving a 1-brane that
is a Dirichlet brane for the captured fundamental
string. Hence we will call it the $(01)b$-string.
 The
tension of the $(01)b$-string equals  
$t_{(01)b} = T_{D2} R_6
= R_6 (T_{A1})^{3/2}/g_A$.
\foot{
Note that there is no Kaluza-Klein tower of particles
since a fundamental string cannot end on the N5 brane. }

Since the $(01)b$-string is a Dirichlet
string for the $(10)b$ string  the
coupling follows from:
\eqn\cmpcup{
t_{(01)b} =  {1 \over  g_b} (t_{(10)b}) \quad. }

Now $(M_{\rm Planck}^{(9)})^7={(T_{1A})^4R_6\over g_A^2}$
so we  take a limit:
\eqn\dcblk{
\eqalign{
(M_{\rm Planck}^{(9)})^2 / t_{(10)b}
=  ( {T_{A1}^{1/2} R_6 \over  g_A^2} ) ^{2/7}
&
\rightarrow \infty \cr
(M_{\rm Planck}^{(9)})^2 / t_{(01)b}
= ( { g_A^{3} \over (R_{6}^{5} T_{A1}^{5/2}) }  ) ^{1/7}
&
\rightarrow \infty\cr} }

The solution of these conditions is 
easily expressed in Taub-NUT variables:
we need only take   $V$ to infinity.
Indeed,    the first expression above is equal to
$(V/g_B^2)^{2/7}$, while the second is
equal to $(V^2 g_B^{3})^{1/7}$.
Note that this is simply the limit in which both coupling constant
$g_{IIA}$ and the radius of the transversal circle $R_6$
  go to zero like $1/V$.

The case $R_6=\infty$  considered in \sls\ seems to
be infinitely far away from the case we are dealing with.
In this case D-brane in the little
string theory is infinitely heavy. 

Similarly to the previous case one can show that
after $T$ duality in the sixth direction
strings on the brane become captured IIB (1,0)
and (0,1) strings. The captured fundamental
string is still a fundamental string after the
$T$-duality,
while wrapped $D2$ brane becomes the captured
$D1$- brane, i.e. the IIB (0,1) string.

\subsec{The vectormultiplet picture: Definition of the
$IIa$-theory from the $IIB$ $D5$-brane}

Now let us consider the IIA string with a wrapped 
$D5$ brane: 

\eqn\lttlea{
IIB:  (\overline{R_1,R_2,R_3,R_4,R_5})_{k,D5},R_6 ; g_B , T_{B1}   
}

The relevant excitations  are the
instanton string\foot{also known as a bound state
of the $D1$-brane with k $D5$-branes} in the low-energy $5+1$
$U(k)$ super-Yang-Mills with tension $t_{a1} = T_{1B}/g_B$,
 together
with the 2-branes and 0-branes obtained by wrapping  
the $D3$ brane and $(1,0)$ IIB string around the sixth
direction. These branes have tensions $t_{a2}=T_{1B}^2 R_6/g_B$
 and
$t_{a0}=T_{1B}R_6$.
 These 0-branes and 2-branes are Dirichlet branes for
the instantonic string and one can easily check
that  $t_{a2}={1 \over  g_a} t_{a1}^{3/2} $
and $t_{a0}={1 \over g_a} t_{a1}^{1/2} $
for $$g_a= 1/(R_6 T_{1B}^{1/2}g^{1/2})$$.

The limit we want is: $g_a$ is fixed, while
\eqn\plnk{
{(M_{\rm Planck}^{(9)})^2 \over t_{1a} }= \bigl(
{(T_{1B})^4 R_6\over g_B^2}\bigr)^{2/7}
{g_B \over  T_{B1}}
  \rightarrow \infty
}
If we pass to the Taub-NUT variables
for $TN5[A]$ , we easily get
 $g_a=g_{IIA}$, while
the left hand side of expression \plnk
is simply given by  $V$.

>From \twoa\ we see that in this limit the  
radius of the transversal circle decreases as
$1/V$ while the coupling constant and tension
grow like $V$.

As in all other cases we will show that
after dualization to the Taub-NUT picture
strings and branes become captured bulk
strings and branes. However, 
 in order to see this we must first go
 to the $S$-dual picture.

\subsec{The vectormultiplet picture: 
Definition of the $IIa$ theory from 
the IIB NS5 brane}

The $S$-dual vectormultiplet picture starts from the 
background: 
\eqn\littlap{
IIB:  (\overline{R_1,R_2,R_3,R_4,R_5})_{k,NS},R_6 ; g_B , T_{B1}  }

This background is $S$-dual to
the IIB $D5$- brane background,
so it is quite obvious that the little
string in this case is still an instantonic string,
(i.e. a captured fundamental string),
with   tension $t_{a1}=T_{1B}$,
while the $0$- and $2$- branes come from
$D1$- and $D3$-branes wrapped arownd the sixth dimension.

The tension of the captured fundamental string
is $t_{a1}=T_{IIB}$, while
the tensions of the branes are
\eqn\tensone{
t_{a0}=T_{IIB}R_6/g_B , }
and
\eqn\tenstwo{
t_{a2}=T_{IIB}^{2} R_6/g_B
}
so the little string coupling constant
follows from 
\eqn\smllstrcp{
t_{a0}=T_{IIB}^{1/2}/g_a
}
and equals
\eqn\smllcpli{
g_a= (T_{IIB}^{1/2} R_6)^{-1} g_B\quad .
}

The condition of decoupling of gravity is
\eqn\dcplgrv{
(T_{IIB}^4 R_6/ g_B^{2})^{2/7}/T_{IIB}
\rightarrow \infty
}
with $g_a$ held fixed.

Once again, it is easy to show that in
Taub-NUT variables $g_a=g_{IIA}$,
and expression \dcplgrv\  behaves like
\eqn\tnvrsin{
(T_{IIB}^{1/2} R_6/g_B^{2})^{2/7}=
(V/g_{IIA}^2)^{2/7}
}
Thus, we get decoupling of gravity
for constant tension,
and for transverse radius and coupling constant
decaying like $1/V$.

After $T$ duality
in the sixth direction the wrapped $D1$-
and $D3$-branes
turn $D0$- and $D2$-branes
  captured by the center of Taub-NUT

\subsec{The $6+1$-dimensional picture: definition 
of $m$-theory}

Finally, to define $m$-theory we 
begin with IIA theory with parameters
\eqn\wrpsx{
IIA:  (\overline{R_1,R_2,R_3,R_4,R_5,R_6})_{k,D6} ;
 g_A  , T_{A1}   
}

The $6+1$ SYM coupling gives the tension of the 
instantonic membrane: 
\eqn\sxpli{
t_{m2}   =
{1 \over  g_{6+1}^2} 
= {1 \over  g_A}  T_A^{3/2}
}
There are now monopole 3-branes, and instantonic 
2-branes, {\it but no strings}. 

We wish to decouple bulk gravity. Therefore, 
\eqn\limplk{
{(M_{\rm Planck}^{(10)})^2\over  t_{m2}^{2/3} } = {g_A^{-1/2} T_A  \over 
t_{m2} } = g_A^{1/6} 
\rightarrow \infty
}
Thus, to define $m$-theory we must introduce 
one scale, $t_{m2}$ and take the limit: 

\eqn\defltlm{
\eqalign{
m: \quad R_1,R_2,R_3,R_4,R_5 , R_6; &  
t_{m2}  \qquad\qquad\qquad \cr
\lim_{g_A \rightarrow \infty } 
IIA:  (\overline{R_1,R_2,R_3,R_4,R_5,R_6})_{D6} ;
 g_A  , T_{A1} = (g_A t_{m2})^{2/3}  \cr}
}
So,   $m$-theory is obtained from the theory
on the $D6$ brane when the string coupling constant  
grows while the tension of $D2$-brane is kept fixed.
In Taub-NUT units $g_A=V^{3/2}$,   again are confirming
the Taub-NUT picture.

Because of its origin,
$m$-theory should be a $6+1$-dimensional 
Poincar\'e invariant theory with $16$ real 
supercharges. Its low energy limit is described 
by $6+1$-dimensional SYM theory and has no 
graviton in the spectrum.  It has neutral 
$2$-branes. 
Note that the definition of $IIa$
 is precisely correct for its 
interpretation  as the compactification 
of $m$-theory on a circle of radius $R_6$.

\subsec{Definition of $f$ theory}
 
In the first definition of $IIb$ theory
we recognize     a lower-dimensional 
version of $F$-theory. Because the M5- and N5-
branes do not change under $T$-duality
the above theories are intrinsically 
$5+1$-dimensional. The coupling torus is not 
geometrical in $IIb$ theory.  This is the 
analog of the familiar story of  
``twelve dimensions'' of $F$-theory 
\vafa\vfmr. 

Moreover, it is significant that we have 
derived $f$-theory from a bulk picture of branes. 
Note that since the $M5$ brane is geometrically 
embedded into the 7-torus we conclude that 
{\it the elliptic fibrations of $f$-theory must have a
section.}

Now it is clear how to define the $f$ theory for the
category of theories without gravity:
it is just a theory of 
branes wrapped around the base in an
ellipticaly fibered  manifold.  

\subsec{The buck stops here}

As a check on the self-consistency of 
our story  will now confirm that the M-theory 
5-brane transverse to a two-torus is the 
maximal theory for which we can take the 
little string limit. 
Consider an $M5$ brane perpendicular to a
three-torus. The condition for decoupling 
gravity becomes:  
\eqn\bthree{
 R_i \ell^{-3} << (M^{(6)}_{\rm Planck})^2
\equiv  \biggl({R_1 R_2 R_3\over \ell^9}\biggr)^{1/3},}
which is inconsistent. 
Thus we conclude once again that b- theory is a
maximal in its class.

\subsec{Remarks on the decoupling of gravity}

In the above discussion we have argued that gravity 
decouples when certain ratios of scales go to zero. 
In fact, there are many subtleties which should be 
addressed before the decoupling of the entire bulk 
theory is really established. 
 We will make some brief remarks on this here. 

The absence of radiation in the Taub-NUT picture 
(with $V\rightarrow \infty$) may be heuristically
understood as follows. The low energy modes of the 
$d$-dimensional supergravity multiplet ($d=10,11$)
when compactified
on Taub-NUT  consist of 
the graviton multiplet and the KK states. First consider
the ``pure gravity'' case. The kinetic term for 
these modes (far from the center) is given by 
the $(d-1)$-dimensional Einstein action with Planck 
mass $M_{d-1}=(M_d^{d-2} RV)^{1/(d-3)}$, where 
$M_d$ is the Planck mass of the $d$-dimensional 
theory. The scale of the extended objects captured 
by the Taub-NUT space is $M_d$. As $V \rightarrow\infty$
the ratio $M_d/M_{d-1} \rightarrow 0$, giving suppression 
of ``pure gravitational radiation.''

The suppression of the KK radiation is more 
complicated, but here we can offer the following 
heuristic argument. The states with KK momentum 
have a mass equal to the inverse radius of the 
Taub-NUT. If such a particle moves towards the 
center of the Taub-NUT then, as the radius 
shrinks the mass effectively grows. This results in 
a repulsion of the particle from the center. 
Consequently the wavefunction of these particles 
vanishes at the center and hence they are not radiated
from the brane. Of course, this heuristic argument 
could be improved by a calculation. 

Under $U$ duality (see below) the 
``pure gravitational radiation'' is mapped
to radiation of the same kind. The KK 
radiation takes various forms in various 
pictures. In the $D6$-picture it is radiation 
of $D0$-branes. These are repelled from the 
$D6$-brane \polchinski\ and hence there is 
no radiation of these particles. 

\newsec{From $U$-duality to $u$-duality}

In this section we discuss the duality symmetries 
of the little theories. Our starting point for the 
compactified theories is $M$ theory on $T^7$. 
Therefore the parent theory has 
$U$-duality group 
$\Gamma= E_{7,7}(\IZ) \equiv E_{7,7}(\IR) \cap Sp(56;\IZ)$
\hull. In the TN picture we are decompactifying one 
circle. Therefore, it is manifest that the duality 
group of the little theory should be 
$u=E_{6,6}(\IZ)$. Still, it is instructive to see how 
$u$-duality arises directly from the other 
5-brane pictures described in the previous section. 

\subsec{Reduced $U$-duality}

We are considering $M$-theory 
on tori $T^d=(S^1)^d$ with a
diagonal metric and 
with $A_{MNP}=0$. This defines a subvariety of
the Teichmuller space $E_{7,7}(\IR)/K$, $K=SU(8)$, 
which
we call the reduced subvariety $\CR$. It contains much of
the essential physics. This subvariety is nothing but a
copy of the Cartan torus. We have Iwasawa decomposition
$U = NAK$ and  the maximal split form has
$A \cong (R^*)^7$. Only a subgroup of the $U$ 
duality group $\Gamma$ will
preserve $\CR$. We call this the reduced $U$-duality
group.

The reduced duality group is easily seen to 
be isomorphic to the Weyl group of 
$E_7$. First note that
 $\gamma \cdot AK/K = AK/K$ implies
that $\gamma=ak$ where $a\in A$ and $k A k^{-1}=A$
gives an action of the Weyl group. The reason is 
that  $\gamma\cdot1\in AK \rightarrow \gamma=ak
\rightarrow a^{-1} \gamma = k \in K$.
Moreover $(k A k^{-1}) K = AK$ so
$k A k^{-1}=A$.
When we further take $\gamma\in   E_{7,7}(\IZ)$ then
we are restricted to the Weyl group $W$. 
Thus the reduced moduli space is canonically
$\CM^{red}= W\backslash A$.

It is interesting to identify the physics associated to  
generators of the reduced duality group. 
\foot{This is closely related to recent work of \egk. 
These authors went beyond $d=7$, obtaining fascinating 
results.}
The Weyl group is generated by reflections in simple 
roots. These reflections may be interpreted in terms 
of permutations of radii, $T$-dualities and $S$-dualities. 
In the following we work this out in the $M$ theory frame: 

We first relate the compactification data
   $R_1, \dots R_7$ in the $M$-theory metric to an 
element in $\CM^{red}$. 
Cremmer and Julia \cj\ 
define the $SL(8,\IR)$ matrix:
\eqn\cji{
\eqalign{
S & = p^{-1/4} Diag\{ R_1, \dots, R_7, p\} \cr
p & = R_1 \cdots R_7\cr}
}
and then construct the $E_{7,7}$ matrix by the
embedding into $Sp(56,\IR)$:
\eqn\cjii{
S \rightarrow \pmatrix{ \rho_{28}(S) & 0 \cr 0 & \rho_{28}(S^{tr,-1})\cr}
}
(The extra $E_7$ generators are defined infinitesimally
by rank 4 antisymmetric tensors, or globally by
preserving the quartic invariant in addition to the
symplectic invariant.)

We now define the $SL(8,\IR)$ roots:
$\alpha_1 = e_1 -e_2,\dots , \alpha_7 = e_7-e_8$.
The spinor root extending $SL(8,\IR)$ to
$E_{7,7}$ is then
\eqn\spin{
\alpha_s   = -\half(e_1 + e_2 + e_3 + e_4) + \half(e_5 + e_6 + e_7+ e_8)
}
Decompactification to $M/T^6, M/T^5,\dots $ etc is
represented by {\it removing}, in order,  from
the Dynkin diagram ({\it without} $\alpha_7$!)
$\alpha_1, \alpha_2,\dots, \alpha_6, \alpha_s$.
Note that $\alpha_7=-\theta$ is the negative of the
highest root in the $E_7$ root system.
If we remove $\alpha_6$ then we get the
string group $SL(2,\IR) \times O(6,6;\IZ)$
while the subgroup $SL(7,\IR)$ is geometrically
manifest in $M$ theory.

Many of the duality formulae involving products 
of radii become more transparent in a logarithmic 
basis. An element of the Cartan subalgebra is
\eqn\csa{
S = \exp[\sum_{i=1}^7 t_i \alpha_i ]
= \exp[\sum_{i=1}^8 y_i e_i ]
}
The $y_i$ are related to the radii $R_i= e^{x_i}$
by $y_i=x_i-s/4$, $i=1,\dots, 7$, $y_8=3s/4$, 
$s=\sum_i x_i$. 
We can now easily compute the action of the Weyl-group
reflections $w_{\alpha}$ on the radii for  various 
roots: 

\item{1.} $w_{\alpha_i}$, $i=1,\dots 6$,   are just
transpositions of the radii $R_i \leftrightarrow R_{i+1}$
holding other radii fixed.

\item{2.} $w_{\alpha_7}$ acts via:
\eqn\alphsv{
\eqalign{
R_i' & = R_i (R_{123456})^{-1/3} \qquad i =  1, \dots 6 \cr
R_7' & = R_7 (R_{123456})^{+2/3}\cr}
}

\item{3.} Finally, the reflection $w_{\alpha_s}$ in the
spinor root acts via:
\eqn\alphspin{
\eqalign{
R_i' &= R_i (R_5 R_6 R_7)^{1/3} \qquad i = 1,2,3,4 \cr
R_i' &= R_i (R_5 R_6 R_7)^{-2/3} \qquad i=5,6,7 \cr}
}

The reflection in the spinor root $w_{\alpha_s}$ 
corresponds to an electromagnetic duality in 
a frame with a $D3$ brane. This should not be confused 
with the electromagnetic duality generated by the 
symplectic form $\Omega$ of $Sp(56;\IZ)$. The latter
generates an action of the Weyl group which is
just $x_i \rightarrow - x_i$, ie. $R_i \rightarrow 1/R_i$.
\foot{Note that this is central.
The Weyl group of $E_7$ has a central element
because all the representations are real.}

The natural generators of the
Weyl group of combine to give the $S$ and $T$-dualities
of the 
string point of view. All relations between various 
duality chains can be derived from the Coxeter
relations of the generators of the Weyl group:
$w_i w_j = w_j w_i $ if the roots are orthogonal
$w_i w_j w_i = w_j w_i w_j$ if they are connected
by a line.

\subsec{Relations between the theories following 
from $U$-dualities of the parent theory }

Having defined the little string theories as in
section 3
we can now express relations between them 
following from the standard relations between the 
10 dimensional string theories and M theory. 
\foot{  The Weyl group of $E_7$ has order
$\vert W(E_7) \vert= 2^{10} 3^4 5 7 = 7! (24)^2 = 2903040$
so there are plenty of other pictures we could discuss!}

We consider the system of $k$ M5-branes \smsxi. 
In general they can be separated (but parallel) in the 
5 transverse directions. 
 
This background has several equivalent descriptions
related by the reduced $U$-duality group. The following 
descriptions correspond to the pictures used in 
the previous section to define the little string theories:

Choosing the $6$-cycle to define a $IIA$ theory gives: 
\eqn\smsxii{
IIA:  (\overline{R_1,R_2,R_3,R_4,R_5})_{NS},R_7 ; g_A=(R_6/\ell)^{3/2}, T_{A1} = R_6/\ell^3 
}
 
Applying $T$-duality: 
\eqn\smsxiii{
IIB:  (\overline{R_1,R_2,R_3,R_4,\ell^3/(R_5 R_6) })_{NS},R_7 ; g_B=(R_6/R_5) , T_{B1} = R_6/\ell^3 
}

Applying $S$-duality: 
\eqn\smsxiv{
IIB:  (\overline{R_1,R_2,R_3,R_4,\ell^3/(R_5 R_6) })_{D5},R_7 ; g_B=(R_5/R_6) , T_{B1} = R_5/\ell^3 
}

We can 
also get other wrapped $D6$-brane pictures, for example 
by applying $T_7$ to \smsxiv. 
\eqn\smsxxxii{
IIA:  (\overline{R_1,R_2,R_3,R_4,{\ell^3 \over R_5 R_6} ,{\ell^3 \over R_5 R_7} })_{D6} ;
 g_A={\sqrt{R_5 \ell^3} \over  R_6 R_7}  , T_{A1} = R_5/\ell^3 
}
and so forth.

Now one can check that if we take the limit defining 
the $IIb$ theory from the $M5$ brane: 
\eqn\limag{
R_6 = t_{b10} \ell^3, \quad R_7 = t_{b01} \ell^3 , \quad
\ell \rightarrow 0
} 
then the corresonding limits in 
the other pictures is in 
accord with the intrinsic definitions 
of the little theories given in the previous section. 
For example, we get the following ``little theories'': 

>From \smsxi\smsxii: 
\eqn\limofa{
IIb: \quad R_1,R_2,R_3,R_4,R_5; 
  g_b=t_{b10}/t_{b01}, t_{b10}
}

>From \smsxiii\smsxiv: 
\eqn\limofb{
IIa: \quad R_1,R_2,R_3,R_4,{1 \over  R_5 t_{b10}} ; 
  g_a={(t_{b10}/t_{b01}) \over  R_5 \sqrt{t_{b10}} }, t_{a1}= t_{b10}
}

>From \smsxxxii: 
>From \smsxiii\smsxiv: 
\eqn\limofaa{
m: \quad R_1,R_2,R_3,R_4,{1 \over  R_5 t_{b10}},
{1 \over  R_5 t_{b01} }  ; 
  t_{m2} = t_{b10} t_{b01} R_5
}

\subsec{$u$-duality}

The couplings and tensions of the little theories
 \limofa\limofb\limofaa\ 
are related by the standard rules of $u$-duality. 
The full $u$-duality group is $u = E_{6,6}(\IZ)$.
This is easily established by checking the 
generators identified above.  For 
example, $s$-duality in $IIb$ theory is just 
transposition $R_6 \leftrightarrow R_7$ of the 
parent $M$-theory. Moreover, the 
$S_5 = W(SL(6,\IZ))$
subgroup of $E_{6,6}(\IZ)$ is also obvious from the 
$M$-theory origin. Combining with $T$-dualities of the 
other string theories includes the action of the spinor 
root and fills out $E_{6,6}(\IZ)$.

\newsec{Some excitations in the $m_k,a_k,b_k$ theories}

The most convenient representation for deriving infinite 
towers of BPS excitations in the little string theories 
is that in terms of $D3$ branes.
Consider $k$  $D3$-branes in type IIB theory
(with the coupling constant $g$, and tension $T_{1B}$) that is
perpendicular to the rectangular torus
with radii $R_1,R_2,R_3$.
\eqn\dthreei{
IIB:  (\overline{R_4,R_5,R_6})_{k}, R_1,R_2,R_3 ; g , T_{1B}  }
The limit we are taking corresponds to
$R_{i}^2 T_{1B} \rightarrow 0$, holding $g$ fixed. 

The massive BPS vectormultiplets are $(p,q)$ strings that
begin and end on the $D3$ brane and wrap the $i^{th}$
circle $n_i$ times. They come in adjoint multiplets of 
$U(k)$ and have masses (with background axion field equal 
to zero): 
\eqn\cone{
T_{1B} \sqrt{(p^2 + q^2/g^2 )} \sqrt{ n_{1}^{2}R_{1}^{2} +n_{2}^{2}R_{2}^{2}+ 
n_{3}^{2}R_{3}^{2} ,}
}
where $n_1,n_2,n_3$ are relatively prime integers, and 
$p,q$ are relatively prime integers. The chirality  of the 
the higher dimensional tensormultiplet theory is reflected 
in the condition $0 \leq p \cdot q $.

Using $T$-dualities and the $M$-theoretic interpretation 
one can easily recover the $7$-dimensional adjoint $U(k)$
vectormultiplet and wrapped 2-branes of   $m$ theory.

In order to get the $b$-theory we first use 
$U$-duality to rewrite \dthreei\ as: 
\eqn\dthreeii{
\eqalign{
IIA:  (\overline{R_4,R_5,R_6,{1\over R_1 T_{1B}} })_{k} & ,R_2,R_3 ; g , T_{1B} \cr
M:  (\overline{R_4,R_5,R_6,{1\over R_1 T_{B1}}, {g\over R_1 T_{1B}} } )_{k } & ,R_2,R_3 ;T_{M2}=R_1 T^2_{1B}/g \cr}
 }
Now we are in a position to give the interpretation of
the vector-multiplet masses from the $b$-theory point of view.
In $b_k$-theory we expect to have $k^2$ $SL(2,Z)$ multiplets
of strings 
with tensions 
\eqn\cthree{
T_{M2} \sqrt{n_{2}^{2}R_{2}^{2} + n_{3}^{2}R_{3}^{2}}=
{T^2_{1B} R_1\over g}  \sqrt{n_{2}^{2}R_{2}^{2} + n_{3}^{2}R_{3}^{2}}. 
}
These are the terms with $n_1=0$ in \cone. 

The terms with $n_1=1,n_2=n_3=0$  are the 
KK modes of the selfdual
tensor field compactified on a torus. This 
field produces the same massless vector multiplet as
a six-dimensional massless vectormultiplet,
while it produces only 1/2 of the massive vector
multiplets. This is reflected by the condition
$0 \leq p \cdot q $.

Of course, the absence of the graviton multiplet in 
all dimensions is consistent with all $U$-dualities.

\newsec{Remarks on ``captured strings''}

We have argued that dualities together with 
the decoupling of the bulk theory inevitably 
lead to the little string theories and their 
$m$-theory counterpart. However, in various 
pictures we see that we must have ``captured 
strings,'' on branes, even at weak string couplings. 
Dualities do not explain physically
how this strange phenomenon 
happens. 

One explanation of the string capturing process 
proceeds by studying the decoupling of the Higgs 
and Coulomb branches of a $D1$ or $D2$ brane probe. 
This subject has been extensively discussed recently. 
Some references include \WittCommt\dps\seibhiggs\seibcoul\
\ref\aspentalks{Seminars by J. Maldacena, R. Dijkgraaf, 
A. Hanany, and E. Silverstein at the July 1997 workshop on 
Nonperturbative String Theory at the Aspen Center for 
Physics.}

Consider $k$ coincident $D6$-branes in the limit
when the string coupling constant goes to infinity.
Let us study the theory on a $D2$-brane, parallel to
the collection of $D6$-branes. This is a
$U(1)$ SYM theory with 8 supercharges.
This theory has bosonic fields of three types:
gauge fields $A$,
hypermultiplets $H_i$, $i=1,...,k$, and three 
scalars $\phi_a$-corresponding to the motion
of $D2$-brane perpendicular to $D6$-brane.
(scalars corresponding to the motion of
a $D2$ parallel to $D6$ are irrelevant in this 
 discussion).
 
 If we canonically normalize
 the kinetic energy for the gauge field
 then the bosonic piece of the lagrangian
 reads, schematically:  
\eqn\ando{F^2 + \sum_a(d\phi_a)^2+ |DH_i|^2 + g^2 (\vec D(H))^2 +
 g^2 (\sum_{a} (\phi_a)^2)(\sum_{i} H_i H_{i}^{*}).}
where $\vec D(H)$ are the D-terms giving
 the hyperkahler moment 
map of the ADHM construction of the
 moduli space of $SU(k)$ instantons 
on $\IR^4$ of instanton number 1. 
The one-loop induced Coulomb branch metric is the periodic 
solitonic 5-brane metric. Dualization with respect to 
the compact scalar produces the Taub-NUT geometry.
\foot{We ignore several confusing issues here.}
The Higgs branch $X=0$, $H \neq 0$
in the theory of the 2-brane theory 
describes the captured phase. 

A full understanding of the capturing phenomenon 
should come from analyzing the decoupling of the 
Coulomb and Higgs branches. In \seibhiggs\
this was discussed in terms the supersymmetric
quantum mechanics of the $D0D4$ system
\dkps. This 
can be regarded as a kind of  
   ``minisuperspace approximation.''
\foot{which is known to work well in various 
models of 2D gravity.}

In this context the capturing phenomenon can 
be heuristically understood as follows. 
Taking the $D4$ to fill the $6,7,8,9$ 
hyperplane the  $U(1)$ 
VM on the $D0$-brane now consists of 
 five scalars $\phi_a$
and  4 complex fermions $\lambda$. 
(We take $A_0=0$ gauge). 
The $k$ HM's give complex scalars $(h^i, \tilde h_i)$
and $k$ together with 4 complex fermions $\psi^i$. The 
energy is: 
\eqn\erngy{
\eqalign{
\CH = \sum_{a=1}^5 (\dot \phi_a)^2 + \sum_{i=1}^k \bigl[ \bigl\vert {d \over dt} h^i\bigr\vert^2
+ \bigl\vert {d \over dt} \tilde h_i\bigr\vert^2\bigr]
&
+ g^2 \vec D(h,\tilde h)^2 
 + g^2 \vec \phi^2 \sum_{i=1}^k ( \bigl\vert   h^i\bigr\vert^2
+ \bigl\vert   \tilde h_i\bigr\vert^2 ) \cr
 + g \sum_{i=1}^k \bar \psi_i \phi_a \Gamma^a \psi^i & +  g\sum_{i=1}^k(h^i \lambda \tilde \psi_i + 
 \tilde h_i \lambda  \psi^i + {\rm cplx.\ conj.})  \cr}
}
where we have written the fermion couplings schematically (they are 
determined by the global symmetries). 

In order to study the Higgs-Coulomb
transition will will introduce a highly simplified caricature. 
We  
consider a  quantum mechanical system with 
two real degrees of freedom $H,X$, intended 
the represent the Higgs and Coulomb branch 
degree of freedom, and two corresponding fermionic 
oscillators $\{ a_H, a_H^\dagger\} =1, \{ a_X, a_X^\dagger\} =1$,
respectively. The Hamiltonian is:  
\foot{Note that this is {\it not} a supersymmetric quantum mechanics, but 
is meant as a simplified model of the essential physics of the 
complicated system \erngy.}
\eqn\hamil{\CH=
\half P_{X}^2 + \half P_{H}^{2}+ \half g^2 X^2 H^2  -\half g\vert X\vert a_H^\dagger a_H 
- \half g \vert H \vert a_X^\dagger a_X }
The little string limit is the limit $g^2 \rightarrow \infty$ at fixed 
energy $E$. This is a highly quantum-mechanical regime.

A fixed energy contour of the potential $  g^2 X^2 H^2 = E $ is 
described by four hyperbolas, the regions $X^2, H^2 \rightarrow \infty$
being the regions of escape to Coulomb and Higgs branches. 
In the region $H^2 \ll X^2$ we can 
use a Born-Oppenheimer approximation to write approximate 
wavefunctions. 
\foot{We thank 
J. Harvey for a very helpful discussion on this point. 
Note that we are ignoring subtleties associated with the 
tube metric here. }
An energy eigenfunction is approximated by $\Psi(X,H) \cong 
\Psi_{\rm slow}(X) \Psi_{\rm fast}(H;X)$ where 
\eqn\brnopp{
\eqalign{
\biggl(\half P_H^2 + \half g^2 X^2 H^2 -\half g\vert X\vert a_H^\dagger a_H 
\biggr) \Psi_{\rm fast}(H;X)
& = E_f^{(n)}(X) \Psi_{\rm fast}(H;X)\cr
\biggl(\half P_X^2 + E_f^{(n)}(X) \biggr) \Psi_{\rm slow}(X)
& = E^{n,m}  \Psi_{\rm slow}(X)\cr}
}
Solving for the fast modes we get an energy spectrum 
$E^{(n)}(X) = n g \vert X\vert$, $n=0,1,2,\dots$. 
For $g \rightarrow \infty$ we have a zero energy state at 
$n=0$, as guaranteed by supersymmetry, and a large gap above 
the supersymmetric ground state. Plugging in the states with 
$n>0$ into the equation for the slow modes leads to 
an expression for $\Psi_{\rm slow}(X)$ which
decays like 
$e^{-{2\over 3} g^{1/2} \vert X\vert^{3/2}}
e^{-\half g \vert X\vert H^2}
$ 
and has a discrete spectrum $E^{n,m}\sim c_{n,m} g^{2/3}$
where the $c's$ are positive constants. These states have 
an enormous gap above  the nearly supersymmetric 
$n=0$ states. The latter have 
$E = \half P^2$ and approximate wavefunction
\eqn\coulwave{
\Psi_{\rm Coulomb} \cong e^{i PX} e^{-g \vert X\vert H^2} a_H^\dagger \vert 0\rangle_{H,X}
}
These represent waves travelling on the Coulomb branch. 
This approximation is good for $g \vert X\vert \gg E$, 
that is, $E/g \ll \vert X\vert $ and hence 
 approaches all $X$ values for $g \rightarrow \infty$. 

An analogous discussion holds for waves on the Higgs branch with 
the replacement $H\leftrightarrow X$, leading 
to approximate energy eigenfunctions: 
\eqn\higgswave{
\Psi_{\rm Higgs} \cong e^{i PH} e^{-g \vert H\vert X^2} a_X^\dagger \vert 0\rangle_{H,X}
}
valid for $E/g \ll \vert H\vert $. 

Now we are ready to estimate the transition probability from 
Higgs to Coulomb branches. We can use \coulwave\ or 
\higgswave\ as an excellent approximation except in the 
transition region $\vert X\vert \leq E /g $ 
{\it and} 
$\vert H\vert \leq E /g $. This region has 
area $E^2/g^2$, so, assuming there is no anomalous 
enhancement of the wavefunctions in this region
\foot{such as might occur if there were a boundstate} the 
contribution to the 
transition amplitude from this region must shrink to 
zero for $g\rightarrow \infty$. Moreover, one can 
estimate the overlap integral
\eqn\overlap{
\Biggl\vert \int dH dX \Psi_{\rm Higgs}^* \Psi_{\rm Coulomb}
\Biggr\vert \leq \CO(g^{-2/3})
}
and we finally conclude that the Higgs to Coulomb transition 
is suppressed for $g \rightarrow \infty$. This is the 
trapping phenomenon. 

If $k=1$ the Higgs branch is just a point,
so the heuristic argument becomes much weaker. 
Nevertheless, duality symmetry still predicts the 
strings should be captured. Subtleties associated with 
the distinction between $k=1$ and $k>1$ are discussed in 
\WittHiggs\seibcoul. It would be very interesting 
to do a more careful and exact analysis of the above 
capturing problem. 

\newsec{Applications and Speculations}

\subsec{Future directions and applications}

It is widely appreciated that there are many potential 
applications for the little string theories. 
One of the main potential applications is to the formulation
of the so-called ``M(atrix) theory'' 
approach to defining M-theory in the light cone gauge in 
certain backgrounds \bfss. The reason is that one 
hopes that little string theories will be more tractable 
and have fewer degrees of freedom than their big brothers. 
Another potential application has been discussed in 
\WittQCD. These theories might also be relevant to 
black hole physics \stromvafa\dps.

We would like to mention another application. 
\foot{The work referred to in this paragraph was 
done in collaboration with A. Gerasimov.}
The motivation for the present work was rather different from 
that of most of the recent papers on little string theories. 
We were motivated by a search for analogues of $(d=4,\CN=2)$ 
super-Yang-Mills theories with a {\it compact Coulomb branch}. 
Such theories might lead to very interesting elliptic  
 generalizations of the prepotentials of 
Seiberg-Witten theory. (A trigonometric extension has been 
given in \nnfive.) 
A natural place to search for such theories 
is in the theory of the $D3$ probe in F-theory 
compactification of IIB on $\IP^1$  \senpillow\bds, 
or in the M2-probe transverse to a K3 surface in M-theory 
on K3 \seibir. However, the low energy theory on the probe 
is not defined by any quantum field theory but rather should 
be a low energy limit of 
some version of the little string theories. Indeed, our 
route to discovering little string theories proceeded by 
systematically
simplifying the problem of understanding the D3 probe in 
F-theory. We hope that an understanding of little string 
theories in the simpler context of toroidally compactified
$M$-theory will lead to 
an understanding of the more general class of theories 
with 8 supercharges and a compact Coulomb branch.

\subsec{Speculations}

Finally, we would like to mention two speculations. 

A radical interpretation of the above discussion is that 
7-dimensional $m$-theory and $11$-dimensional $M$-theory 
are different phases of some master theory - call it 
${\bf M}$-theory. In ${\bf M}$-theory even the number 
of spacetime dimensions is not well-defined, and takes 
different values in different phases. Thus, $M$ and $m$ 
theory are simply different phases of one theory, roughly 
analogous to the Coulomb and Higgs branches of SYM theories, 
respectively. Indeed, this analogy becomes precise when 
considering the effective 2+1 dimensional theory of a 
2-brane probe. 

An even more radical speculation posits that there is 
an infinite tower of theories in $2\mod 4$ dimensions of 
which $m$- and $M$- theory are but the first two examples. 
The line of reasoning suggesting this idea is the following: 

The existence of little string theories clarifies some 
 mysterious aspects of $F$-theory \vafa\vfmr.  Some aspects 
of $F$-theory suggest that it is twelve dimensional. However
such an interpretation leads to two problems:

\item{Q1.} Why are there  no Kaluza-Klein modes
coming from   11- and 12 -dimensions?

\item{Q2.}  Why do the elliptically-fibered
manifolds on which we compactify require a section?

The description of b-theory as an  M5 brane
perpendicular to a two-dimensional 
torus suggests the following answers: 

\item{A1.} The 10-dim type IIB theory is not a compactification of 
a 12-dimensional theory on a torus, rather 
it is the theory of a brane transverse to a two-torus, 
embedded in some higher-dimensional spacetime.

\item{A2.} If we assume that 10D spacetime is a brane, we
immediately conclude that the manifold, over which we
``compactify'' - in the sense of $F$-theory, 
needs a section, defined by the position of the brane. 

This viewpoint raises the question to what extent
$M$-theory  is fundamental and  whether the 11-dimensional 
spacetime of $M$ theory is simply a brane in some higher-dimensional
space.\foot{We are grateful to
A. Gerasimov for posing this question and for suggesting 
the following idea.}
One could go on to speculate that   the theory in which
10-dim and 11-dim theories are embedded is  at least 14 dimensional,
and it contains all known branes and a new symmetry,
continuing the sequence:  gauge symmetry, general covariance, .......?

\bigskip
\centerline{\bf Acknowledgements}
\bigskip

We would like to thank Anton Gerasimov for collaboration
in the initial stage of this work and for sharing his ideas.
G.M. would like to extend special thanks to Jeff Harvey and 
Erik Verlinde for extensive discussions. 
We also thank many colleagues including
O. Aharony, T. Banks, T. Damour, J. Harvey, M. Kontsevitch, D. Kutasov, 
E. Martinec, N. Nekrasov, N. Seiberg, S. Shenker, 
E. Silverstein, E. Verlinde, and E.Witten
for  stimulating discussions. 
A. L. would like to thank the LPTHE and especially 
L. Baulieu for hospitality, SS. would like to thank the IHES
for hospitality during his stay in Paris in summer 97 when 
this manuscript was finished. 
G.M. would like to thank CERN and the Aspen Center for Physics 
for hospitality during the completion of this work. 
The research of A.L. is supported by RFFI grant 
950101101, and grant 96-15-96455 for support of 
scientific schools. The research of G. M. 
is supported by DOE grant DE-FG02-92ER40704,
and by a Presidential Young Investigator Award; that of S. S.,
by DOE grant DE-FG02-92ER40704, by NSF CAREER award, by
OJI award from DOE and A. P. Sloan Foundation.

\listrefs

\end

\newsec{Decoupling of gravity and large radius of Taub-NUT}
Here we present some estimations showing
that decoupling  of the bulk gravity really takes
place in m,a,b-theory with the finite coupling constant
 and that 
 in the Taub-NUT picture in corresponds  to very large radius
limit of Taub-NUT.

\subsec{The m-theory and a-theory}
Here story is very simple -
when $V$ of the corresponding Taub-NUT goes to infinity,
in the  $D6$ picture it corresponds to
$g$ going to infinity,
i.e. 2-brane, captured by $D6$ brane becomes
very light.
Explicitely, from the M-theory
point of view the 10-dim Plank
mass equals $(M^9 R_{11})^{1/8}$,
while the scale determined by the captured 
2-brane equals to $T^{1/3}=M$.
Now note, that the limit $g$ going to infinity
is exactly the limit $V$ going to infinity.

$a$-theory  could be obtained by compactification of $m$-theory
so above argument is unchanged (see section 6). 

\subsec{The b-theory}

Let us consider the following  realization of $b$-theory -
$(M5,R_1,R_2)$.  Branes, wrapped around circles lead
to strings  with tensions $M^3 R_1$ and $M^3 R_2$.
Here $M$ is a 11-dimensional Plank mass.
Coupling constant is just the ratio of
tensions $R_1/R_2$. 
We expect that radia are very small, so when we are
interested in radiation from the world-volume of the brane
we are interested in the 9-dimensional reduction of
the 11-dimensional supergravity. 
So, we have to compare the tension of the string on a brane with
the square of the 9-dimensional Plank mass $M_9$, where

\eqn\bzero{(M_9)^7=R_1 R_2 M^9.}

Thus, condition for decoupling gravity from the 
$b$-string theory reads:
\eqn\bone{
M^3 R_i << (R_1 R_2 M^9)^{2/7},
}
i.e.  (if small radia have the same order of magnitude,
that we  denote as $R$):
\eqn\btwo{
M R <<1.
}

Note, that if we try to play the same game
for a configuration $M5$ brane perpendicular to a
tree torus, then the same condition
would read as:
\eqn\bthree{
M^3 R_i << (R_1 R_2 R_3 M^9)^{2/6}= M^3 (R_1 R_2 R_3)^{1/3},}

i.e. it is contradictory.
Thus we conclude once again that b- theory is a
maximal in its class (as we expected).

Now, we will pass to IIB picture using  
relation: $M/T_2=IIB/T_1$
and get that the radius of
$T_1$ (that equals to the radius of corresponding Taub-NUT ) equals to:
\eqn\bfour{
V^2=R_{TNUT}^2 T_1 = 1/(M^3 R_1 R_2^2),}

i.e. it really goes to infinity in the limit where gravity decouples.

\newsec{Phenomenology of $m,a,b$ }

Here we will show that theories $m,a,b$ really have
the properties announced in the section "Outline". Here we will study
the theories with $N=1$. We will comment on cases with higher $N$ at the end
of the section 5.

\subsec{Branes in $m_1,a_1,b_1$ theories}

1. $m_1$-theory: from the interpretation
of $m_1$-theory as a theory on the worldvolume of the
sigle
D6 brane it is clear that  $m_1$ theory has a
2-brane - it is just bound states of
$D2$ brane with the $D6$ brane(see 
"Appendix".)

{From the point of view of
the low-energy 7-dimensional
theories this is just  an instantonic
2-brane in $U(1)$ of the zero size.
  
{From the point of view of 3-dimensional theory on
a 2-brane the capturing is just the Higgs phase
of the $U(1)$ gauge theory. 
This Higgs phase consists of a point
degenerate with the endpoint of the Coulomb branch.
The existence of such phase is a nontrivial problem in the 
field theory, but it is predicted by dualities
(like a bound state of two $D0$ branes, see [Witten,July 97]
for a discussion on that point.

{From the same $D6$-picture it is clear that
there are no other branes in the theory.

This confirms prediction that TNUT
in $M$ theory really captures 
the membrane.

2. Let us study
$a_1$-theory from the point of view of $NS(B)$ brane.
 It
has a string that is just a fundamental string captured by $NS(B)$
brane
(like in case 1. this string could be considered
as an abelian instantonic string),
and 0 and 2 branes (
The ending of $D4$-brane on
the NS(B)-brane is discarded here due to
the low codimension of resulting brane) 
coming from $D1$ and $D3$-branes starting
and ending  
on the $D5$-brane and wrapping around the perpendicular
circle.
Thus, string in $a_1$ theory could end
on 0 and 2-brane,
i.e. the latter branes are $d$-branes (small $d$ 
means in small theory).
 
By $T$-duality in the perpendicular circle 
to reach the Taub-NUT picture we
find that string and  $d$-branes are nothing but
string and
$D$-branes of type IIA theory captured
by Taub-NUT.

3. The pattern of branes $b_1$-theory is 
maximally transparent in the $M5$-theory:
they are just coming from the 
$M$-theory 2-brane starting end ending
on the $M5$-brane, so there  form 
an $SL(2,Z)$ multiplet.

In order to see that they could end on
each other we pass from the $M$-theory picture to
IIA picture using M-theory - type IIA correspondence.
One of the wrapped branes turns into
a fundamental string captured by NS(A)-brane
while another turns into a $D2$-brane 
starting and ending on NS(A)-brane and    
wrapping a perpendicular circle,
thus captured fundamental string can end
on the wrapped $D2$ brane that looks like a string
from the $b_1$-theory point of view.

After performing $T$-duality one easily recognizes
that these strings are just $SL(2,Z)$ multiplet of strings 
of type IIB captured by Taub-NUT.